\documentstyle[12pt]{article}

\newcommand{\stk}[1]{\stackrel{*}{\overline}}

\begin{document}

\begin{center}

\vfill

{\large {\bf {The Supersymmetric Stueckelberg Mass and Overcoming\\ the

Fayet-Iliopoulos Mechanism for Breaking Supersymmetry} }}

\end{center}
\vfill
\begin{center}
S.V. Kuzmin\\ D.G.C. McKeon\\ Department of Applied Mathematics\\ University
of Western Ontario\\ London\\ CANADA\\ N6A 5B7
\end{center}
\vfill
email: DGMCKEO2@.UWO.CA \\ Tel: (519)661-2111, ext. 88789\\ Fax:
(519)661-3523 \eject

\section{Abstract}

Gauge invariant generation of mass for a supersymmetric $U(1)$ vector field
through use of a chiral Stueckelberg superfield is considered. When a
Fayet-Iliopoulos $D$ term is also present, no breaking of supersymmetry ever
occurs so long as the Stueckelberg mass is not zero. A moduli space in which
gauge symmetry is spontaneously broken arises in this case.

\section{Introduction}

The Stueckelberg mechanism for generating a mass for a $U(1)$ vector field
is well understood [1]; it also has a supersymmetric generalization in which
the Stueckelberg field is a chiral superfield [2]. In this case, both the
photon and photino field develop a degenerate mass.

The breaking of a supersymmetry in a supersymmetric $U(1)$ gauge theory can
be accomplished through the presence of a so-called Fayet-Iliopoulos $D$
term [3].

In this paper, we consider what happens when a $U(1)$ supersymmetric gauge
theory is supplemented by both a Stueckelberg chiral superfield and a
Fayet-Iliopoulos $D$ terms. It is demonstrated that if the Stueckelberg mass
is non-zero, then supersymmetry remains unbroken for any value of the
Fayet-Iliopoulos parameter $\xi$, and that gauge symmetry is broken in a
moduli space characterized by the vacuum expectation value of the scalar
matter field.

\section{A $U(1)$ Supersymmetric Model}

We begin with a real $U(1)$ superfield $V=V^{*}$, the associated field
strength $W_\alpha =\overline{D}^2D_\alpha V$, and a chiral matter
superfield $\Phi $. (The conventions are those of [4].) The Lagrangian 
$$
{\cal {L}}_{CL}=\frac 1{32}\left( W^\alpha W_\alpha \right) _F+\left( \Phi
^{\dagger }e^{2gV}\Phi \right) _D\eqno(1) 
$$
possesses the gauge invariance 
$$
V\rightarrow V^{\prime }=V+i\left( \Lambda -\Lambda ^{\dagger }\right) 
\eqno(2a) 
$$
$$
\Phi \rightarrow \Phi ^{\prime }=e^{-2ig\Lambda }\Phi \eqno(2b) 
$$
where $\Lambda $ is a chiral (gauge) superfield. One can supplement ${\cal {L
}}_{CL}$ of eq. (1) by the Fayet-Iliopoulos term [3] 
$$
\left. {\cal {L}}_{F1}=\xi V\right| _D\eqno(3) 
$$
without breaking gauge symmetry. If $\xi g<0$, then spontaneous breaking of
gauge symmetry occurs, leaving supersymmetry unbroken, while if $\xi g>0$
supersymmetry is broken and gauge symmetry is unbroken.

One can also introduce a chiral superfield $S$ that acts as a Stueckelberg
field [2]. It is possible then to have a gauge invariant mass term 
$$
\left. {\cal {L}}_M = m^2\left[V + \frac{i}{m} \left(S -
S^\dagger\right)\right]^2\right|_D\eqno(4)
$$
provided 
$$
S \rightarrow S^\prime = S - m\Lambda\eqno(5)
$$
when the transformation of eq. (2) occurs.

In the Wess-Zumino gauge in which $V$ becomes 
$$
V=\theta \sigma ^\mu \overline{\theta }V_\mu +i\theta \,\theta \,\overline{
\theta }\,\overline{\lambda }-i\overline{\theta }\,\overline{\theta }
\,\theta \lambda +\frac 12\theta \theta \overline{\theta }\overline{\theta }D
\eqno(6) 
$$
and the matter field $\Phi $ can be expanded as 
$$
\Phi =\phi +\sqrt{2}\,\theta \psi +\theta \theta F+i\partial _\mu \phi
\theta \sigma ^\mu \overline{\theta }-\frac i{\sqrt{2}}\,\theta \theta
\partial _\mu \psi \sigma ^\mu \overline{\theta }--\frac 14\partial ^2\phi
\theta \theta \overline{\theta }\,\overline{\theta }\eqno(7a) 
$$
$$
\Phi ^{\dagger }=\phi ^{\dagger }+\sqrt{2}\,\overline{\theta }\,\overline{
\psi }+\overline{\theta }\,\overline{\theta }F^{\dagger }-i\partial _\mu
\phi ^{\dagger }\theta \sigma ^\mu \overline{\theta }+\frac i{\sqrt{2}}
\overline{\theta }\,\overline{\theta }\theta \sigma ^\mu \partial _\mu 
\overline{\psi }-\frac 14\partial ^2\phi ^{\dagger }\theta \theta \overline{
\theta }\,\overline{\theta },\eqno(7b) 
$$
it follows that 
$$
{\cal {L}}_{CL}+{\cal {L}}_{FI}=-\frac 14V_{\mu \nu }V^{\mu \nu }+i\lambda
\sigma ^\mu \partial _\mu \overline{\lambda }-\frac 14V_{\mu \nu
}^{\;\;\;*}V^{\mu \nu }+\frac 12D^2\nonumber 
$$
$$
+\left( D_\mu \phi \right) ^{\dagger }\left( D^\mu \phi \right) +i\psi
\sigma ^\mu D_\mu ^{\dagger }\overline{\psi }\eqno(8) 
$$
$$
+F^{\dagger }F+i\sqrt{2}g\left( \phi ^{\dagger }\psi \lambda -\phi \overline{
\psi }\overline{\lambda }\right) \nonumber 
$$
$$
+g\left( \phi ^{\dagger }\phi D\right) +\xi D\;,\nonumber 
$$
$D_\mu =\partial _\mu +igV_\mu $ and $V_{\mu \nu }=\partial _\mu V_\nu
-\partial _\nu V_\mu $. If we now parameterize the Stueckelberg field as 
$$
S=\left( \frac{A-iB}2\right) +\sqrt{2}\,\theta \chi +\theta \theta
F_S+i\partial _\mu \left( \frac{A-iB}2\right) \theta \sigma ^\mu \overline{
\theta }\nonumber 
$$
$$
-\frac i{\sqrt{2}}\,\theta \theta \partial _\mu \chi \sigma ^\mu \overline{
\theta }-\frac 14\,\partial ^2\left( \frac{A-iB}2\right) \theta \theta 
\overline{\theta }\,\overline{\theta }\eqno(9) 
$$
then in the Wess Zumino gauge, ${\cal {L}}_M$ in (4) becomes 
$$
{\cal {L}}_M=\frac{m^2}2\left( V_\mu -\frac 1m\partial _\mu A\right)
^2-\frac 12B\partial ^2B-i\left( \chi \sigma ^\mu \partial _\mu \overline{
\chi }-\partial _\mu \chi \sigma ^\mu \overline{\chi }\right) \nonumber 
$$
$$
+2F_S^{\dagger }F_S+mBD-\sqrt{2}\,m\left( \overline{\chi }\overline{\lambda }
+\chi \lambda \right) .\eqno(10) 
$$
The residual gauge invariance in (8) and (10) 
$$
V_\mu \rightarrow V_\mu +\partial _\mu \Lambda \eqno(11a) 
$$
$$
A\rightarrow A+m\Lambda \eqno(11b) 
$$
$$
\phi \rightarrow e^{-ig\Lambda }\phi \eqno(11c) 
$$
$$
\psi \rightarrow e^{-ig\Lambda }\psi \eqno(11d) 
$$
can be broken by a so-called ``$U$ gauge'', 
$$
A=0\eqno(12) 
$$
or by an ``$R$ gauge'' with a gauge fixing Lagrangian 
$$
{\cal {L}}_{GF}=-\frac 1{2\alpha }\left[ \partial \cdot V+\alpha mA\right]
^2\;.\eqno(13) 
$$
In the former case, $A$ is completely eliminated and $V^\mu $ is a massive
vector with the longitudinal polarization present while in the latter case, $
A$ just decouples from $V$ and the longitudinal contribution of $V_\mu $
vanishes as $\alpha $ goes to zero. If one were to neglect the axial anomaly
that occurs, the renormalizability of the model is apparent in the $R$ gauge
of eq. (13).

The potential from (8) and (10) is given by 
$$
V=-\left\{ \frac 12D^2+F^{\dagger }F+2F_S^{\dagger }F_S+g\phi ^{\dagger
}\phi D+\xi D+mBD\right\} \;.\eqno(14) 
$$
Using the equation of motion for $D$, $F$ and $F_S$ this becomes 
$$
V=\frac 12\left( g\phi ^{\dagger }\phi +\xi +mB\right) ^2\;.\eqno(15) 
$$
It is evident that provided $m\neq 0$, the minimum of $V$ is zero
irrespective of the values of $g$ and $\xi $. If $\phi $ has a vacuum
expectation value of $\phi _0$, then the vacuum expectation value of $B$ is 
$$
B_0=-\frac{\xi +g\phi _0^{\dagger }\phi _0}m\eqno(16) 
$$
in order to minimize $V$. Since this minimum occurs at $V=0$, supersymmetry
is unbroken. There is thus a ``moduli space'' for the scalar fields $\phi $, 
$\phi ^{\dagger }$ and $B$ in which supersymmetry is unbroken; if $m\neq 0$
then supersymmetry is in fact never broken. Only if $m=0$ can the
Fayet-Iliopoulos mechanism for breaking of supersymmetry be operative.

If now $f$ and $b$ are the quantum fluctuations of $\phi $ and $B$
respectively about the background field, so that 
$$
\phi _0=\phi _0^{\dagger }=h\eqno(17a) 
$$
$$
\phi =h+f,\;\;\;\;\;\phi ^{\dagger }=h+f^{\dagger }\eqno(17b) 
$$
$$
B=-\left( \frac{\xi +gh^2}m\right) +b\;,\eqno(17c) 
$$
then we find from (8), (10), (13) and (15) that in the Wess-Zumino $R$ gauge
the component field form of our Lagrangian will have a term bilinear in $f$
and $V^\mu $. To eliminate this cross term, we modify the gauge fixing of
eq. (13) in a way suggested by 't Hooft [5,6] so that 
$$
{\cal {L}}_{GF}=-\frac 1{2\alpha }\Big[\partial \cdot V+\alpha \left(
mA+2ihf\right) \Big]\left[ \partial \cdot V+\alpha \left( mA-2ihf^{\dagger
}\right) \right] .\eqno(18) 
$$
Together, (8) (10), (15) and (18) leave us with the total Lagrangian 
$$
{\cal {L}}=-\frac 12\left[ \left( \partial _\mu V_\nu \right) \left(
\partial ^\mu V^\nu \right) -\left( 1-\frac 1\alpha \right) (\partial \cdot
V)^2\right] +\frac 12\left( m^2+2g^2h^2\right) V_\mu V^\mu \nonumber 
$$
$$
+\left( D_\mu F\right) ^{\dagger }\left( D^\mu f\right) +\frac 12\left(
\partial _\mu A\right) ^2-\frac 12b\partial ^2b\eqno(19) 
$$
$$
-\frac 12\alpha \left[ mA+2ihf\right] \left[ mA-2ihf^{\dagger }\right] 
\nonumber 
$$
$$
-\frac 12\left[ gh\left( f+f^{\dagger }\right) +mb\right] ^2--gf^{\dagger
}f\left[ gh\left( f+f^{\dagger }\right) +mb\right] \nonumber 
$$
$$
-\frac 12\left( gf^{\dagger }f\right) ^2+i\lambda \sigma \cdot \partial 
\overline{\lambda }-i\left( \chi \sigma \cdot \partial \overline{\chi }
-\partial \chi \cdot \sigma \overline{\chi }\right) \nonumber 
$$
$$
+i\psi \sigma \cdot D^{\dagger }\overline{\psi }-\sqrt{2}\,m\left( \overline{
\chi }\overline{\lambda }+\chi \lambda \right) \nonumber 
$$
$$
+i\sqrt{2}gh\left( \psi \lambda -\overline{\psi }\overline{\lambda }\right)
+i\sqrt{2}g\left( f^{\dagger }\psi \lambda -f\overline{\psi }\,\overline{
\lambda }\right) \nonumber 
$$
once the auxiliary fields have been eliminated. For all values of the vacuum
expectation value $h$, this model does not have spontaneously broken
supersymmetry provided $m\neq 0$. The vector field has a mass $\left(
m^2+2g^2h^2\right)^{\frac{1}{2}}$.

\section{Discussion}

We have considered a $U(1)$ gauge model in which a real vector superfield $V$
has been coupled to a chiral matter superfield $\Phi $. This has been
supplemented by a Fayet-Iliopoulos term and, through the use of an
additional chiral superfield $S$, a Stueckelberg mass term. It is found that
the model always has a supersymmetric ground state, and that there is a
moduli space for the vacuum expectation value $h$ of the scalar component of 
$\Phi $ much as there is in non-Abelian $N=2$ supersymmetric models. If $
h\neq 0$, gauge symmetry is spontaneously broken.

One could also consider what happens when more then one chiral matter
superfield couples to the vector superfield as in the Wess Zumino model of
QED [7] with a Fayet Iliopoulos $D$-term [3]. We have found that if a
Stueckelberg mass term is also present, it is again not possible to have
spontaneous breaking of supersymmetry provided the Stueckelberg mass is
non-zero. However, in this case, if there is a gauge invariant mass term for
the matter fields, the vacuum state is unique and this vacuum state, the
expectation value of the scalar matter fields is zero.

There are several avenues of investigation that suggest themselves. One
might consider coupling directly the Stueckelberg superfield $S$ to the
matter superfield $\Phi $ in a gauge invariant fashion. This has been done
in [8]; the results indicate the possibility of devising a model in which
supersymmetry is broken and the vector field remains massless. It is also
tempting to consider a non-Abelian generalization of the model considered in
the preceding section. Possibly the divergences normally encountered in
conventional non-Abelian Stueckelberg models [9] are mitigated by
supersymmetry.

\section{Acknowledgments}

NSERC supplied financial support. R. and D. MacKenzie gave a useful
suggestion.


\begin{thebibliography}{9}
\bibitem{1}  E.C.G. Stueckelberg, {\it Helv. Phys. Acta.} {\bf 11}, 225
(1938).
\bibitem{2}  J.L. Buchbinder and S.M. Kuzenko, {\it Ideas and Methods of
Supersymmetry and Supergravity }(Institute of Physics, Bristol,1998), Ch.
3.4.6; \\ R. Delbourgo, {\it J. Phys. G.} {\bf 1}, 800 (1975).
\bibitem{3}  P. Fayet and J. Iliopoulos, {\it Phys. Lett.} {\bf 51B}, 461
(1974); \\ P. Fayet, {\it Nuovo Cimento} {\bf 31A}, 626 (1975).
\bibitem{4}  D. Balin and A. Love, {\it Supersymmetric Gauge Field Theory
and String Theory} (IOP Publishing, Bristol, 1994).
\bibitem{5}  G. 't Hooft, {\it Nucl. Phys.} {\bf B35}, 167 (1971); \\ K.
Fujikawa, B.W. Lee and A.I. Sanda, {\it Phys. Rev.} {\bf D6}, 2923 (1972).
\bibitem{6}  S.V. Kuzmin and D.G.C. McKeon, {\it Mod. Phys. Lett.} {\bf A 16}
, 747 (2001).
\bibitem{7}  J.Wess and B.Zumino, {\it Nucl.Phys.} {\bf B78}, 1 (1974).
\bibitem{8}  S.V. Kuzmin and D.G.C. McKeon, in {\it MRST 2002, AIP
Conference Proceedings, Vol. 646}, eds. V. Elias, R.J. Epp, and R.C. Myers (
Melvill, New York, 2002), pp.111-116.
\bibitem{9}  M. Veltman, {\it Nucl. Phys.} {\bf B21}, 288 (1970).
\end{thebibliography}
\end{document}